\documentclass[12pt]{article} 

\usepackage[german,english]{babel}

\usepackage{amscd,amsmath}
\usepackage {amsfonts,apalike}
\usepackage{epsfig,graphicx,graphics}

\usepackage{cancel}

\usepackage{undertilde}
\usepackage[normalem]{ulem} 

\usepackage{epsfig,graphicx,graphics}

\usepackage{cancel}
\DeclareGraphicsExtensions{.pdf,.png,.jpg}

\title{First-Class Constraints, Gauge Transformations,  de-Ockhamization, and Triviality:  Replies to Critics, Or, How (Not) to Get a Gauge Transformation from a Second-Class Primary Constraint}

 \author{J. Brian Pitts \\Programme in Philosophy,  University of Lincoln, \\ Faculty of Philosophy,  University of Cambridge,   and \\ Department of Philosophy, University of South Carolina  } %

 \date{\today}

\begin{document}

\maketitle

\begin{abstract} 

Recently two pairs of authors have aimed to vindicate the longstanding conventional claim that a first-class constraint generates a gauge transformation in typical gauge theories such as electromagnetism, Yang-Mills and General Relativity, in response to the Lagrangian-equivalent reforming tradition, in particular Pitts, \emph{Annals of Physics} 2014.  Both pairs emphasize the coherence of the extended Hamiltonian formalism against what they take to be core ideas in Pitts 2014, but both overlook Pitts 2014's sensitivity to ways that one might rescue the claim in question, including an additive redefinition of the electrostatic potential.  Hence the  bulk of the paper is best interpreted as arguing that the longstanding claim about separate first-class constraints is \emph{either false or trivial}---de-Ockhamization (using more when less suffices by splitting one quantity into the sum of two) being trivial.  Unfortunately section 9 of Pitts 2014, a primarily verbal argument that plays no role in other works, is refuted. 

Pooley and Wallace's inverse Legendre transformation to de-Ockhamized electromagnetism with an additively redefined electrostatic potential, however, opens the door to a precisely analogous calculation introducing a photon mass, which shows that a \emph{second-class primary} constraint generates a gauge transformation in the exactly same sense---a \emph{reductio ad absurdum} of the claim that a first-class constraint generates a gauge transformation and a second-class constraint does not.  Gauge freedom by de-Ockhamization does not require any constraints at all, first-class or second-class, because any dynamical variable in any Lagrangian can be de-Ockhamized into exhibiting trivial additive artificial gauge freedom by splitting one quantity into the sum of two.  Physically interesting gauge freedom, however, is typically generated by a tuned sum of first-class constraints.

\end{abstract}

keywords:  gauge freedom, observables, constrained Hamiltonian dynamics, determinism 


\section{Introduction}

Recently two teams of philosophers of physics have undertaken to defend a traditional ``orthodox''  claim that a first-class constraint \emph{alone} (not as part of a team) generates a gauge transformation in electromagnetism and analogous theories (such as Yang-Mills and General Relativity), whether in detail \cite{PooleyWallacePittsGauge} or in a footnote \cite{GomesButterfieldMaxwellGauge}, in response to my 2014 \emph{Annals of Physics} paper \cite{FirstClassNotGaugeEM}.  This debate is interesting not primarily in terms of light shed on electromagnetism, which is well understood, but for what it suggests about General Relativity, which has many mathematical analogies to Maxwell's theory, but where calculations are much harder and supposedly change is missing, especially from the supposedly fundamental physical quantities, ``observables'':  the ``problem of time'' (\emph{e.g.,} \cite{KucharCanadian92,KucharCanonical93,IshamPrima,IshamQuestionTime,AndersonProblemofTime,GRChangeNoKilling,ObservablesEquivalentCQG,BergmannObservables}). 
 My own work has been  inspired by and partly built upon the Lagrangian-equivalent reforming literature under way since the 1980s, especially works by various combinations of Shepley, Pons, Salisbury, and Sundermeyer, as the lengthy bibliography shows. Generally my work  has aimed  aimed to evaluate  traditional lore and largely verbal arguments with calculations where possible, seeking reflective equilibrium between general principles and independently understood examples,  inspired by Sundermeyer \cite{Sundermeyer} and Goodman \cite[p. 64]{Goodman}.
Sundermeyer's  treatment of constrained Hamiltonian dynamics expects the general theory to match prior understanding of the examples (which is less common than one might expect).  
Goodman discusses how to line up generalities (whether relating to inductive logic or the meanings of words) by mutual adjustment of candidate generalities and uncontroversial examples, a process that came to be known as reflective equilibrium (due to John Rawls).  
  This example-tested approach is  vindicated even in the breach, ironically, by my section 9's making a largely verbal and invalid argument, followed by Pooley and Wallace's supplying the necessary mathematical corrective; mathematical arguments generally are better than verbal ones.  While it is certainly unfortunate that section 9 of Pitts 2014 is erroneous, critics seem not to notice the distinction between section 9 and the rest of the paper.  Coming to grips with the argument of Pooley and Wallace will turn out to provide the materials for a new argument further showing that the separate first-class constraints doctrine is \emph{either false or trivial} (as Pitts 2014 outside section 9 argued):  it will appear how a \emph{second-class primary} constraint in massive Proca electromagnetism generates a gauge transformation in just the same sense that Pooley and Wallace employ for the separate first-class constraints in Maxwell's electromagnetism.  But everyone agrees that a second-class constraint does not generate a gauge transformation.  \emph{Felix culpa.}

Pooley and Wallace reasonably say that
\begin{quote} 
[i]t would indeed be surprising (to say the least) if the conventional view
could be shown to be wrong via an easy calculation that somehow had never
been done explicitly. In this paper we shall argue that Pitts is mistaken. \cite{PooleyWallacePittsGauge} 
\end{quote} 
Fortunately I do not claim that the calculation in question had never been done, but that ``such a test apparently
has not been made before, at least not completely and successfully, and has rarely been attempted.''  
That description still seems to me largely correct:  the calculation has not been done nearly so often as it should have been, and its point has been largely missed.  
But the surprising possibility was arguably already realized previously by Pons's calculation \cite{PonsDirac}, though it is perhaps not as easy as mine. 
Mine did not compare two trajectories and so had no cancellation, whereas Pons, like Dirac and those who follow him  \cite{HenneauxTeitelboim}, compared two trajectories, consequently had some terms cancel out, and, in Pons's case (not Dirac's), worked sufficiently far ahead  to achieve a mathematically correct result.  Pooley and Wallace ought to identify some mistake in Pons's work also if they are vindicating the conventional claim about separate first-class constraints.  

Given that a  reforming program has been under way since the 1980s, it is hardly surprising if the conventional view does have some serious flaw, so one should assess one's probabilities based on more of the available evidence.   It would not be very surprising either if a new argument or a new way of looking at old ideas exposed the flaw in the conventional view in a clearer way. Pons's work \cite{PonsDirac} was perhaps overly modestly entitled ``On Dirac's Incomplete Analysis of Gauge Transformations.''  Pons showed that, contrary to Dirac, a first-class constraint does not generate a gauge transformation in the sense of preserving Hamilton's equations, because Dirac worked only to first order in time and one sees the need for a tuned sum of first-class constraints, primary and secondary\ldots, at higher order.  A glance at the citation history for this paper by Pons will show that it did not quickly make the impact that it deserved, certainly not by 2013, probably due to its  modest title, its publication in a philosophy of physics journal, and  perhaps its location in the history and philosophy of physics section the physics ArXiV (albeit with cross-listings in gr-qc, hep-th, and math-ph).    
Understatement seems not to be working if a conventional wisdom and a mathematical critique can coexist in the literature for decades (since the 1980s) without proponents of the former even noticing and feeling any need to respond.  Hence it seemed appropriate to make  a clear statement of what is at stake, recognizing the separate \emph{vs.} combined constraints issue to be a watershed in the problem of time in canonical quantum gravity.  I see no need to preserve a paradox if it can be resolved instead, at least classically.  Understatement and deference are not the style among analytic philosophers:  the \emph{Library of Living Philosophers} series honors distinguished philosophers by having other philosophers evaluate and criticize their work, not simply list and celebrate it or contribute something vaguely relevant to a \emph{Festschrift}.

While my 2014 paper frequently makes disjunctive (``or'') remarks such as ``a first-class constraint does either something permitted but pointless (a position-dependent field redefinition) or something bad (spoiling Gauss's law),'' Pooley and Wallace neither notice these disjunctive claims nor make any effort to address the triviality worry.  This sort of disjunctive claim implies that most of my paper does not assert that there is no way of rescuing the separate first-class constraints claim (except section 9).  While the paper title is quite explicit, one can rarely stuff a precise thesis into a title.  I expected that readers would judge an unmotivated position-dependent field redefinition to be too trivial a proposal to take seriously as a form of gauge freedom. To see why, imagine a world much like ours but in which constrained Hamiltonian dynamics has not been invented, and then a non-famous person (not Dirac), or a famous person subject to blind peer review, proposes reformulating electromagnetism by replacing the electrostatic potential $V$ by $V$ plus some arbitrary function (as in the Pooley-Wallace Lagrangian), and then claims to have found additional gauge freedom in electromagnetism.  It is difficult to imagine such a claim being considered interesting. 
 Pooley and Wallace have not explained why their proposal is not trivial, nor even recognized the worry.  They have, however, clarified to what claims proponents of the  conventional wisdom are logically committed.    

At issue is what one is trying to achieve---at a more specific level than simply ``truth'' and perhaps ``justice.''  My goal is to understand electromagnetism and other gauge theories, address the problem of time in General Relativity, and possibly contribute thereby in some fashion to quantum gravity and quantum field theory by clarifying the classical field theoretic foundations, where the mathematics and the interpretation should be comparatively straightforward.  There are other possible goals, such as maintaining a Kuhnian paradigm, or defending a thesis of the inerrancy of Dirac or Henneaux and Teitelboim or perhaps one's teachers in graduate school, or refuting bluntly stated error, perhaps. (I was introduced to  constrained Hamiltonian dynamics in graduate school by Larry Shepley, a proponent of Hamiltonian-Lagrangian equivalence, in case it matters.)

To give a  philosophical analogy, consider the logical problem of evil \cite{LogicalProblemofEvil}, which was advanced  by ``atheologian'' J. L. Mackie, to the effect that there is a \emph{logical contradiction} between the usual claims of an all-powerful, all-knowing, all-good God and the existence of evil.  This logical problem  has been to a substantial degree given up in favor of evidence-based  arguments from evil against theism in the wake of Alvin Plantinga's distinction between a theodicy and a defense and provision of a fairly widely received defense (at least among proponents of libertarian freedom).  In this context, a theodicy tries to give the reason that God permitted evil, or at least a plausible and important reason for such, whereas a defense merely aims to show that there is \emph{no logical contradiction} between the usual  attributes ascribed to God and the existence of evil \cite[p. 28]{Plantinga}. A defense does not need to be true or even plausible.  The precedent is giving a model of non-Euclidean geometry \cite[p. 25]{Plantinga}.  Continuing the analogy, my objection to the separate first-class constraints doctrine is akin to an atheologian's presenting the problem of evil (not necessarily in deductive form) as an objection against theistic belief.  Do Pooley and Wallace need the analog of a theodicy---an interesting and plausible answer---or merely the analog of a defense---a strategy, perhaps contrived and \emph{ad hoc} if necessary, simply for avoiding contradiction---for the separate first-class constraints doctrine?  Again it depends on what one is  trying to do.  If they aimed merely to refute section 9 of my paper, which claims that the separate first-class constraints doctrine (more specifically the extended Hamiltonian) is  wrong (not either wrong or trivial),  somewhat akin to the logical problem of evil, then their work suffices as a defense:  Pitts's 2014 section 9 is refuted.  If, on the other hand, Pooley and Wallace aimed to vindicate the separate first-class constraints doctrine, construed as a long (1950s+) and widely held  interesting and plausible characterization of first-class constraints that is not also true of some second-class constraints, then they need something more like the analog of a theodicy:  they need to find not a triviality but an interesting gauge freedom from the separate first-class constraints in electromagnetism that is absent from theories generally regarded as having no gauge freedom.  Unfortunately they do not notice the difference, and so present a contrived defense-like argument that does in fact refute my section 9, as though it were a substantive theodicy-like argument that refutes Pitts 2014  as a whole and vindicates the  interesting claim about first-class constraints widely received since the 1950s.


\section{Trivial Gauge Freedom for Any Dynamical Variable in Any Theory by De-Ockhamization}

The Pooley-Wallace inverse Legendre transformation from the extended Hamiltonian to a Lagrangian with a redefined electrostatic potential is offered as vindicating the longstanding ``orthodox'' (as they put it) claim that a first-class constraint generates a gauge transformation in electromagnetism.  The fact that one arrives at a de-Ockhamized Lagrangian might give one pause, however.  Has one really learned something about electromagnetism and its gauge freedom?  Or has one merely been treated to an over-interpreted instance of a trick, de-Ockhamization (splitting one quantity into two), that one could apply to any dynamical quantity in any theory, one with first-class constraints, one with second-class constraints, one with both, or even one with no constraints?

To start to answer this question, one can take a Lagrangian with no constraints, perhaps the simplest Lagrangian imaginable, a free particle of mass $1$ in one dimension:  $ L_f = \frac{1}{2} \dot{x}^2.$  Presumably everyone will agree that this theory has no gauge freedom.  Neither does it have any constraints.  Trivially one can give a Hamiltonian formulation:  $H_f = \frac{1}{2} p^2.$    Installing gauge freedom  and a first-class constraint into this `theory' will help us to see how uninteresting the gauge freedom from additive field redefinitions is, given that it can be installed into any dynamical variable in any theory admitting a Lagrangian formulation.  Instead let us use a Lagrangian
$$ L = \frac{1}{2} (\stackrel{.}{x} + \stackrel{.}{y})^2.$$
One can either treat $\dot{y}$ as an arbitrary function or treat $y$ as a dynamical variable.  Let us do each in succession. Treating $\dot{y}$ as an arbitrary function implies that there is an Euler-Lagrange equation only for $x$.  It is: 
$$ - \frac{\partial L}{\partial x} + \frac{d}{dt} \frac{\partial L}{\partial \dot{x} } = 0 + \frac{d}{dt}(\dot{x} + \dot{y}) = 0.$$
This is, of course, just the equation for a free particle of mass $1$ in one dimension when one adds an arbitrary prescribed velocity to $\dot{x},$ or perhaps rather, when one takes the velocity and breaks it into two pieces, one a dynamical variable and one an externally prescribed function.  Obviously this move is not very interesting, at least not without some further motivation. Clearly one can take any dynamical quantity in any theory and de-Ockhamize it in this fashion. One could de-Ockhamize by addition, subtraction, multiplication or division.  While unmotivated and uninteresting, this de-Ockhamized free particle is similar in important respects to the de-Ockhamized electromagnetism of Pooley and Wallace, in which the electrostatic potential $V$ is de-Ockhamized.

Most philosophers of science dislike \emph{ad hoc} hypotheses and think that scientists also dislike them---but is constrained Hamiltonian dynamics an exception?  According to Samuel Schindler, ``[i]t is widely agreed amongst scientists and philosophers alike that a good hypothesis ought not to be ad hoc.'' \cite{SchindlerAdHocCoherentist} 
According to Karl Popper, recalling conclusions already drawn by him in 1919-20, 
\begin{quote} 
[some] genuinely testable theories, when found to be false, are
still upheld by their admirers---for example by introducing \emph{ad hoc} some
auxiliary assumption, or by re-interpreting the theory \emph{ad hoc} in such a
way that it escapes refutation. Such a procedure is always possible, but
it rescues the theory from refutation only at the price of destroying, or
at least lowering, its scientific status. (I later described such a rescuing
operations as a `\emph{conventionalist twist}' or a `\emph{conventionalist strategm}'.)  \cite[p. 48]{PopperCR}
\end{quote}
Imre Lakatos developed a taxonomy of \emph{ad hoc}ness \cite[p. 180]{LakatosChangesInductiveLogic} \cite{AdHocEnthymemePark}.  
Clark Glymour's critique of Hans Reichenbach's conventionalist philosophy of geometry, while not fully convincing regarding its immediate target \cite{SliBimGRG,BenMenahemConventionalism,DuerrBenMenahemConvention},   
nicely captures the uninteresting character of  unmotivated de-Ockhamization.
\begin{quote} 
Suppose \ldots that a bright and articulate student named
Hans one day announces that he has an alternative theory which
is absolutely as good as Newtonian theory, and there is no
reason to prefer Newton's theory to his. According to his
theory, there are two distinct quantities, gorce and morce; the
sum of gorce and morce acts exactly as Newtonian force does.
Thus the sum of the gorce and morce acting on a body is equal
to the mass of the body times its acceleration, and so on. Hans
demands to know why there is not quite as much reason to
believe his theory as to believe Newton's. What do you answer?

I should tell him something like this. His theory is merely
an extension of Newton's. If he admits that an algebraic
combination of quantities is a quantity, then his theory is
committed to the existence of a quantity, the sum of gorce and
morce, which has all of the features of Newtonian force, and for
which there is exactly the evidence there is for Newtonian
forces. But in addition his theory claims that this quantity is the
sum of two distinct quantities, gorce and morce. However, there
is no evidence at all for this additional hypothesis, and
Newton's theory is therefore to be preferred. That is roughly
what I should say, and I believe it is a natural thing to say; but
then I am, I admit, in the grip of a philosophical theory.

The gorce plus morce theory is obtained by replacing
``force'' whereever it occurs in Newtonian hypotheses by ``gorce
plus morce'', and by further claiming that gorce and morce are
distinct quantities neither of which is always zero. 
 \cite{GlymourEpist} \end{quote} 
One does not need to be committed to Glymour's bootstrap theory of confirmation in order to feel that his dim view of unmotivated de-Ockhamizations is basically right. 
But the conventional view that first-class constraints generate gauge transformations with associated sympathy for the extended Hamiltonian seems to be at least implicitly an exception to the claim that philosophers and scientists dislike \emph{ad hoc} theories.  Part of the merit of Pooley and Wallace's work is making clear how, at least in some key examples, the first-class constraint view commits one to valuing de-Ockhamization as an illuminating procedure, rather than a trivial and \emph{ad hoc} trick.

The de-Ockhamized free particle is slightly more interesting if one treats $y$ as a dynamical quantity and employs the mathematical formalism of constrained Hamiltonian dynamics (see, \emph{e.g.}, \cite{Sundermeyer}).  
 One can define canonical momenta:
$p = \frac{\partial L}{\partial \dot{x}} = \dot{x} + \dot{y},$  $\pi = \frac{\partial L}{\partial \dot{y}} = \dot{x} + \dot{y}.$ Letting $y$ bear the burden of the constraint, one has the primary constraint $$\pi - p:$$  the two momenta are equal.  One can define the canonical Hamiltonian using the primary constraint to eliminate $\dot{y}$:
$$ H_c = p \dot{x} + \pi \dot{y} - L =  \frac{1}{2} p^2.$$  The total or primary Hamiltonian, which gives Hamiltonian equations equivalent to the Euler-Lagrange equations $\stackrel{..}{x} + \stackrel{..}{y}$, is $$H_p = H_c + \dot{y} (\pi - p).$$ (Some authors like to keep the unsolved-for velocity(s) and express the primary constraint(s) as constraining some specific momentum/a \cite[pp. 92, 93]{SudarshanMukunda} \cite{CastellaniGaugeGenerator}, as I do here, while others prefer to introduce an arbitrary function that is dynamically forced to equal that velocity or at least be related to it. Keeping the velocities is leaner and works for most interesting theories.)  One  sees that the time evolution from $H_p$ dynamically preserves the primary constraint:  $$\{ \pi - p, H_p\} = \{\pi-p, \frac{1}{2}  p^2 + \dot{y}(\pi - p) \} =0.$$  Why this expression vanishes depends on what one makes of the Poisson bracket of a velocity (assuming that one has a velocity as opposed to an arbitrary function that eventually equals the velocity):  the Anderson-Bergmann rule \cite{AndersonBergmann,ObservablesEquivalentCQG} implies $$ \{\pi-p,  \dot{y}(\pi - p) \} = (\pi - p) \frac{d}{dt} \{ \pi-p,y \} = (\pi -p) \frac{d(-1)}{dt}=0$$ even without invoking the vanishing of the coefficient $\pi - p,$ that is, the primary constraint. Another view is that the Poisson bracket of a velocity is undefined, but harmless because multiplied by $0$ \cite[p. 97]{SudarshanMukunda}.  The vanishing coefficient suffices to give the result at least on the primary constraint surface even without appeal to the Anderson-Bergmann rule or if one uses an arbitrary function instead of a velocity.  So there are many paths that lead to the right answer.  
One can find the generalized Hamilton's equations:  \begin{eqnarray}  \dot{x} = \{x, H_p \} = p - \dot{y}, \nonumber \\
						\dot{y} = \{ y, H_p\} = \dot{y}, \nonumber\\
						\dot{p} = \{ p, H_p\} = - \frac{\partial H_p}{\partial x}=0, \nonumber \\
						\dot{\pi} = \{\pi, H_p\} = -\frac{\partial H_p}{\partial y}=0.  \end{eqnarray}
One sees that the equation for $y$ is empty and that a combination of the other equations yields $\stackrel{..}{x} = 0 - \stackrel{..}{y},$ equivalent to the Euler-Lagrange equation for $x$ (or, for that matter, for $y$).  Because there is only one constraint, its first-class nature follows from the anti-symmetry of the Poisson bracket.  While typically the gauge generator is a tuned sum of first-class constraints of various generations (primary, secondary, perhaps higher) \cite{AndersonBergmann,CastellaniGaugeGenerator,PonsDirac}, in this case the chain has only one link and the primary first-class constraint generates a gauge transformation: 
\begin{eqnarray} 
\{ x, \xi(t) (\pi - p) \} = - \xi(t), \nonumber \\
\{ y, \xi(t) (\pi - p) \} = \xi(t), \nonumber \\
\{ p, \xi(t) (\pi - p) = 0, \nonumber \\
\{ \pi, \xi(t) (\pi - p) \} = 0.
\end{eqnarray} 
As expected from a glance at the de-Ockhamized Lagrangian, the gauge freedom is that when $x$ zigs in an arbitrary way, $y$ zags in an equal and opposite fashion, leaving invariant the quantity $x+y$:   $\{ x+y, \xi(t) (\pi - p) \} = 0,$ much as one might expect given that $x+y$ is the de-Ockhamized replacement for the original $x$.   
Clearly we have learned  nothing about the free particle and rather little about gauge freedom in this exercise.\footnote{We might gather data for the definition of ``observables'', however.  Above  the principle of equivalent observables for equivalent theory formulations, which includes  the requirement that observables are not altered by gauge-fixing \cite{ObservablesEquivalentCQG} has been used. 
When applied to massive electromagnetism \cite{ObservablesEquivalentCQG}, that principle vindicates the gauge generator-based definition of observables \cite{PonsSalisburySundermeyerFolklore} over the separate first-class constraints definition.  
 When applied to massive gravity, it yields a Lie derivative (covariance rather than invariance) because the gauge freedom compares different space-time points \cite{ObservablesEquivalentCQG,ObservablesLSEFoP}. The formal similarity to General Relativity coupled to four scalar fields is overwhelming, so the same definition should be applied in GR, yielding local observables that change and vary spatially.  In the case of a  free particle, the `position' coordinate(s) $x$ and $x+y$ are compared at the same time, so no Lie derivative is needed and one gets invariance of observables, not covariance as in massive gravity and GR. Having the electromagnetic field strength be an observable for Einstein-Maxwell also requires that observables be only covariant (changing by a Lie derivative, not by $0$) under the gauge generator for coordinate transformations \cite{ObservablesEinsteinMaxwellFoP}. 
 Thanks to Oliver Pooley for past discussions about whether invariance \emph{vs.} covariance readily lines up with internal \emph{vs.} external symmetries in defining ``observables''. }   We have gotten a little practice with a simple example where one already knows the answers, which is useful.  One thing that has become clear is that by de-Ockhamization, in particular an additive field redefinition, we can install artificial gauge freedom at will and, if we are so inclined, can generate such modest and physically meaningless gauge freedom with a first-class primary constraint.  That was achieved in an initially unconstrained theory.  We should not be surprised if one can install physically meaningless gauge freedom artificially in a theory of a different sort---whether Maxwell's electromagnetism (where a tuned sum of first-class primary and secondary constraints generates gauge freedom identical to the usual non-Hamiltonian gauge freedom) as Pooley and Wallace and many others do, or in Proca's massive electromagnetism (!) (where the constraints are second-class, so there is not naturally any gauge freedom), as will appear below.

 If a first-class constraint generates a gauge transformation only in non-standard senses such as de-Ockhamization or a canonical transformation  that works just as well for a second-class primary constraint as for a first-class primary constraint \cite[p. 403]{FirstClassNotGaugeEM}, then the late 1950s+ conventional wisdom about separate first-class constraints is hardly vindicated.  Neither de-Ockhamization nor canonical transformations require first-class constraints or indeed any constraints at all; one can deploy such tricks on a free particle, as has been done above, to `find' (that is, install) gauge freedom anywhere.  There exist interesting cases of artificial gauge freedom \cite{Ruegg,PittsArtificial,FrancoisArtificialSubstantialGauge}, but unmotivated de-Ockhamization is not among them.


\section{Can One Assume Consistency in Interpreting Views and Assessing Novelty?}

According to Pooley and Wallace, ``the uncontroversial moral of Pitts's central calculations''  is that ``the constraint-transformed histories fail to solve Hamilton's equations, and hence are not symmetries of the Hamiltonian formalism.''  While I am pleased that Pooley and Wallace accept that moral, I do not think that it is uncontroversial.  Indeed its denial is considered a key result by Dirac \cite[pp. 20-24]{DiracLQM}: he starts with the total Hamiltonian and then convinces himself that the primary constraints, if first-class, generate gauge transformations.  He conjectures that secondary first-class constraints do so as well.  These supposed discoveries about the total Hamiltonian then motivate making this newly discovered gauge freedom manifest by extending the Hamiltonian.  Henneaux and Teitelboim present basically the same argument and  conclusion \cite[pp. 16-18]{HenneauxTeitelboim}, as the 2014 paper discussed in detail.  While one can indeed find Henneaux and Teitelboim making use of both the extended Hamiltonian and the total Hamiltonian in chapter 3 (as Pooley and Wallace observe), giving different gauge symmetries for the corresponding actions,  that does not change the fact that in chapter 1 they followed Dirac in claiming to discover using the total Hamiltonian a larger gauge freedom, which then made it appropriate---indeed they use the word ``must''---to extend the Hamiltonian.  If one agrees that the constraint-transformed histories do not solve Hamilton's equations (for the total Hamiltonian), then the argument of Dirac and of Henneaux and Teitelboim is unintelligible: there is no reason to accept that first-class constraints separately generate gauge transformations, and thus there is no reason to extend the Hamiltonian.  Pooley and Wallace do not explain how they motivate the separate first-class constraints doctrine and the extended Hamiltonian while agreeing with me that ``the constraint-transformed histories fail to solve Hamilton's equations, and hence are not symmetries of the Hamiltonian formalism.''  It seems that they are left with a self-consistent scheme that preserves a verbal formula ``first-class constraints generate gauge transformations,'' but without the original argument or any evident replacement.  Why would one want to preserve the verbal formula if one agrees that the usual argument for it is wrong and does not provide a better one?

In many contexts a judgment of interpretive charity can be assumed, so that an author who says $X$ in one place can be assumed not to be committed also to not-$X$; unfortunately experience with the constrained Hamiltonian dynamics literature shows that one cannot  infer from an author's saying $X$ in one place that he or she is not also committed to not-$X$.  To maintain such a judgment of charity despite such hostile data would involve holding degrees of belief that diverge from the available relative frequencies.  
Dirac convinced himself and many others that a first-class constraint generates a gauge transformation, supposedly a discovered feature of a constrained Hamiltonian theories,  not an arbitrary theoretical choice of de-Ockhamization among constrained Hamiltonian formulations.  He seemed not to realize that he was thereby committed to saying that one \emph{must} extend the Hamiltonian because otherwise that gauge freedom that really is present, is not gauge freedom in his earlier total/primary Hamiltonian formalism.  This seems to be a widely shared inconsistency, thinking that the extended Hamiltonian is a good option, when it should be mandatory if it is good and avoided if it is bad.  Henneaux and Teitelboim might well realize in chapter 3 that separate first-class constraints do not generate gauge symmetries of the total/primary Hamiltonian, but how then does one account for their following Dirac's argument in chapter 1?  Indeed they announce that this idea that first-class constraints generate gauge transformations is a fundamental principle in their work \cite[p. 18]{HenneauxTeitelboim}.  Peter Bergmann spent decades holding incompatible beliefs about gauge transformations (such as involving the gauge generator \cite{AndersonBergmann} and separate first-class constraints \cite{BergmannSchiller}) and also about observables, as been shown recently at length \cite{BergmannObservables}.

Neither have philosophers been immune to inconsistency.  John Earman articulated the  view that a first-class constraint generates a gauge transformation, and also presented  Pons's work (with various collaborators) on the gauge generator (a tuned some of first class constraints) in the same paper, with nary a worry evident that either Pons's gauge generator is too narrow (failing to generate some of the gauge transformations that really exist) or separately smeared first-class constraints are too broad (generating transformations that are not really gauge transformations) \cite{EarmanOde}.  
Pooley and Wallace have also held that first-class constraints really do generate gauge transformations---not simply that it is coherent to take them to do so as a matter of arbitrary theoretical choice by using the extended Hamiltonian---thus not noticing that they are committed, like Dirac, to the claim that the mathematical rule-based Lagrangian-equivalent total/primary Hamiltonian formalism is wrong because it lacks symmetries that the relevant Hamiltonian theory formulations objectively have. There might be some logical inconsistency between my erroneous section 9 and the rest of the 2014 paper, though careful statements of ``false'' and ``either false or trivial'' are logically compatible.  

Regrettable as the lack of consistency in the literature may be, at any rate one sees that one cannot infer that the lesson of my core calculations is uncontroversial, for it is frequently denied as a core principle or key result \cite[p. 21]{DiracLQM} \cite[p. 18]{HenneauxTeitelboim}.  In any case perhaps this sad history suggests why  it is important to seek logical clarity in this field.  The alternative, remaining content with contradictions in the literature for decades, leaves unclear what scholarship is for.  In a literature  tainted by contradictions, saying something true and novel, $X$, counts for less, and recognizing its full significance in excluding everything implying not-$X$, counts for more than is typically the case.


\section{False, or Merely Either False or Trivial?}

Apart from section 9, the  paper \cite{FirstClassNotGaugeEM} repeatedly offers  triviality, not only falsehood, as a possible characterization of the claim about separate first-class constraints---a qualification overlooked by both Pooley and Wallace and by Gomes and Butterfield.  That is because Pitts 2014 was frequently  aware (outside section 9) of the possibility of de-Ockhamization of the electrostatic potential. Section 2 says that ``a first-class constraint does either something permitted but pointless (a position-dependent field redefinition) or something bad (spoiling Gauss's law).''  
Section 10 on canonical transformations takes a disjunctive view systematically:  the idea that a first-class constraint generates a gauge transformation in electromagnetism is either false or trivial. Given that exegesis of the 2014 paper is a key issue, albeit one quite neglected, I take the liberty of quoting my earlier self repeatedly to show how extensively the disjunctive view was presented in section 10.  
``Physical equivalence is preserved, but only by losing some of the original fields' meanings by making an awkward position-dependent field redefinition, it will turn out.''
\begin{quote} 
\emph{By using the full apparatus of a canonical transformation and keeping track of the fact that $Q^0$ is no longer
(up to a sign) the electromagnetic scalar potential} as $q^0$ is, one can resolve the contradiction about vanishing
\emph{vs.} nonvanishing divergence of the canonical momentum \emph{vis-a-vis} the electric field. Such reinterpretation,
which strips the new canonical coordinates of some of their usual physical meaning and
replaces them with a pointlessly indirect substitute, though mathematically permitted, \emph{is certainly not
what people usually intend when they say that a first-class constraint generates a gauge transformation.} \end{quote} 
``To avoid the contradiction of a physics-preserving transformation that changes the physics, one can and must
re-work the connection between $Q^0$ and $A_0$, as shown.''
``The electric field $\vec{E}$, which is an observable by any reasonable standard, is no longer specified simply by
(derivatives) of the new canonical coordinates $Q$, but requires the arbitrary smearing function $\epsilon$ used
in making the change of field variables also. That is permissible but hardly illuminating.''  
``As the generator of a canonical transformation, a first-class constraint does not generate a gauge transformation in
massless electromagnetism any more than a second-class constraint generates a gauge transformation
in massive electromagnetism. Both generate permissible but pointless field redefinitions.''  This appeal to parity of reasoning with massive electromagnetism foreshadows a result derived below, namely, that  some second-class constraints generate gauge transformations by the Pooley-Wallace standard.  
Finally, ``each isolated first-class constraint offers a choice of two bad things (one disastrous, one
merely awkward): it can either destroy the field equations if used directly in Poisson brackets, or
generate a confusing change of physical meaning of the variables as the generator of a canonical
transformation.''  
While it is certainly regrettable that Pitts 2014 section 9 contains key claims that are mistaken---especially the claim that the extended Hamiltonian is empirically inadequate and therefore false---the substantial presence in section 10 of true claims subtly in tension with  the errors localized in section 9 clearly problematizes Pooley and Wallace's claim to find anything problematic for the paper as a whole or the broader systematic questions involved. 
The extent to which section 9 was aberrant becomes clearer:  it contains claims in tension  with claims elsewhere in the paper, and it relies heavily on verbal as opposed to mathematical arguments.  Checking verbal arguments with calculations remains a good idea  even when I fail to follow it but Pooley and Wallace do.  
  Once one sees the disjunctive claim ``false or trivial'' offered as a characterization of the separate first-class constraints doctrine, it becomes appropriate to attend to the question of triviality, but 
 Pooley and Wallace have not attempted to explain why their de-Ockhamized electromagnetism's further gauge freedom is interesting rather than \emph{ad hoc} and trivial. 


\section{Gauge Transformation of States vs. Histories}

In light of the equations $p = \frac{\partial L}{\partial \dot{q} }$ and $\dot{q} = \frac{\partial H}{\partial p},$ the correspondence between the Lagrangian and Hamiltonian formalisms is essentially diachronic.  Whereas Lagrangian states are a bit thick in time, involving both $q$ and $\dot{q}$ (typically) and so requiring an arbitrarily small but nonzero temporal extension for their definition, Hamiltonian states are infinitely thin, involving only $q$ and $p$ at one instant.  Because $p$ only acquires physical meaning diachronically from $\dot{q} = \frac{\partial H}{\partial p},$ it is far from clear that one can speak of a physical state at an instant in phase space.  One has a mathematical state and a Cauchy problem. Diachronically the succession of mathematical states has physical significance.  $q$ has physical significance (assuming that one has not made canonical transformations to destroy the significance of $q$, and also allowing for gauge freedom that might reduce the physical meaning of $q$), but $(q,p)$ at $t$ is not a physical state 
apart from $\dot{q} = \frac{\partial H}{\partial p}$ because $p$ does not have physical meaning at an instant. 
One can safely call $(q,p)$ at $t$ a physical state as long as one does not imperil $\dot{q} = \frac{\partial H}{\partial p}$; the extended Hamiltonian indeed imperils that equation and thus deprives $(q,p)$ at $t$ of the right to be called a physical state.  Consequently I attach no significance to any  criterion for gauge equivalence of Hamiltonian states, such as the Pooley-Wallace criterion involving separate first-class constraints, that competes with  claims about histories.  Pons sees Dirac as thinking of gauge transformations as acting on states in phase space \cite{PonsDirac}, so these points might have broad application.  Consequently any reasoning about physical Hamiltonian states must give way to reasoning about histories, in case of conflict.  Supposing that the phase space Gauss law holds at a single instant (if one takes up electromagnetism as an example), there is no fact of the matter about whether the real Gauss law involving the electric field built from potentials holds at that instant.  The canonical momenta $\vec{\pi}$ conjugate to $\vec{A}$ are simply auxiliary fields in the canonical Lagrangian $p \dot{q} -H$---that is, fields that can be algebraically eliminated using their Euler-Lagrange equations $\dot{q} - \frac{\partial H}{\partial p}=0$ \cite{PonsSubstituting,HenneauxAuxiliary}.  Such surplus structure cannot have fundamental significance if one takes the original Lagrangian as the proper starting place, which is surely the majority view in theoretical physics, or even as \emph{a} proper starting place.  (There are also arguments in favor of using Lagrangians \cite{HojmanShepley,CurielLagrangian}.) That is not to say that one can measure $\dot{q}$ by itself in electromagnetism; one instead measures a gauge-invariant quantity built from $\dot{q}$ and $q$, the electric field, along with the magnetic field built from $q$. The electric field is equal to $\vec{\pi}$ using Hamilton's equations (give or take a sign depending on one's conventions), but that does not imply that one can measure $\vec{\pi}$ if Hamilton's equations are themselves imperiled by extending the Hamiltonian.



\section{Do Second-Class Primary Constraints Also Generate Gauge Transformations? }

Pooley and Wallace appear  not to distinguish  between two very different claims:  (1) that a first-class constraint really does generate a gauge transformation in electromagnetism, as a \emph{fact} about Maxwell's theory (among others) in Hamiltonian form that Dirac \emph{discovered}, and (2) that one can devise a coherent formalism (motivated or not) according to which one can preserve the traditional claim that a first-class constraint generates a gauge transformation as a matter of theoretical choice.  Pitts 2014 unfortunately denied claim (2) in section 9, while more or less affirming it as a triviality in section 10 among other places.  Pooley and Wallace aptly argue for claim (2), especially  by performing an inverse Legendre transformation to what one might call an `extended' Lagrangian, a Lagrangian with a de-Ockhamized electrostatic scalar potential.  They  do little or nothing to support claim (1), however.

After arriving at the extended Hamiltonian (their equation 27), Pooley and Wallace calculate the resulting Hamilton's equations, which are a modification of the Maxwell equations with the gradient of an arbitrary function added in the case of equation 28 for $\dot{\vec{A}}.$  They immediately add that
\begin{quote}  
[w]e can swiftly verify that the transformations generated by $O_t[f,g]$ [their expression that takes arbitrary combinations of the \emph{separately} smeared primary and secondary constraints] are symmetries of \emph{these} equations for arbitrary $f,$ $g.$
 \end{quote}
But this is too quick. Proceeding more slowly, one sees that one can find symmetries  if one is willing to redefine arbitrary functions in the equations---something that one does not have to do with the uncontroversial Lagrangian-equivalent gauge transformations that transform the $4$-vector potential by a $4$-dimensional gradient.  Hence the extended Hamiltonian requires a lowering of standards in order to call the resulting transformations gauge transformations.  Perhaps one can live with this, as long as these lowered standards do not commit one to finding gauge freedom in places that by common consent do not involve gauge freedom.  Unfortunately for the `orthodoxy' that Pooley and Wallace aim to preserve, just such a tragedy occurs, as I now show using Proca's massive electromagnetism.  

One gets Proca's theory of massive electromagnetism---commonly described as involving massive photons, though the quantum words are used even for the classical theory---by adding a ``photon mass'' term $ - \frac{1}{2} m^2 (\vec{A}^2 - V^2)$  \cite[pp. 597-601]{Jackson} \cite[pp. 183-186]{Sundermeyer}. Massive electromagnetism approaches massless (Maxwell) as $m \rightarrow 0,$ both classically and in quantum field theory   \cite{BelinfanteProca,Glauber,BoulwareYM,SlavnovFaddeev,GoldhaberNieto2009,UnderdeterminationPhoton}.
Using $c=1$ and $\hbar=1$ one can convert a mass into an inverse length (tied to the Compton wavelength).  
 Taking the Legendre transformation, one flips the sign of the mass term, so the Hamiltonian (whether canonical or total/primary) sprouts a term $ \frac{1}{2} m^2  (\vec{A}^2 - V^2)$.
The canonical momentum $\pi_0$ is unaffected, so the primary constraint $C_0 = \pi_0$ is just as in Maxwell's theory.  The dynamics shows that dynamically preserving the primary constraint yields a modified Gauss-like law 
$$ C_1^m = \nabla \cdot \vec{\pi} - \rho + m^2 V.$$
One therefore has the total or primary Hamiltonian (corresponding to equations (9), (11) and (12) of Pooley and Wallace) after dropping a boundary term along the way:
\begin{eqnarray} 
H^m =  \int d^3 x \left[ \frac{1}{2} (\vec{\pi}^2 + \vec{B}^2) +  \lambda \pi_0 + \vec{\pi} \cdot \nabla V + (\rho V + \vec{A} \cdot \vec{J})  + \frac{1}{2} m^2(\vec{A}^2 - V^2) \right] = \nonumber \\
= \int d^3x \left[ \frac{1}{2} (\vec{\pi}^2 + \vec{B}^2) +  \lambda C_0 - V \nabla \cdot \vec{\pi}  + (\rho V + \vec{A} \cdot \vec{J})  + \frac{1}{2} m^2(\vec{A}^2 - V^2) \right] = \nonumber \\
\int d^3x \left[ \frac{1}{2} (\vec{\pi}^2 + \vec{B}^2) +  \lambda C_0 - V ( \nabla \cdot \vec{\pi}  -\rho  + m^2 V)    + \vec{A} \cdot \vec{J}  + \frac{1}{2} m^2(\vec{A}^2 + V^2) \right] \nonumber \\
= \int d^3x \left[ \frac{1}{2} (\vec{\pi}^2 + \vec{B}^2) +  \lambda C_0 - V C_1^m    + \vec{A} \cdot \vec{J}  + \frac{1}{2} m^2(\vec{A}^2 + V^2) \right]
 \end{eqnarray}  
where 
$$ C_1^m =  \nabla \cdot \vec{\pi}  -\rho  + m^2 V = 0$$ 
is the modified massive Gauss law in phase space.  Note that the primary constraint $C_0= \pi_0$ is exactly as in Maxwell's electromagnetism, a point that will prove crucial later.  

This massive Hamiltonian, provided for reference, will not be the starting point for the de-Ockhamized massive Hamiltonian to be found later, because it is not clear what, if anything, one can  do to a Hamiltonian with second-class constraints in order to `find' some kind of gauge freedom; indeed one would expect the answer to be ``nothing'' because it is generally agreed that second-class constraints do not generate gauge transformations.  At some point there was a discussion of whether second-class constraints might contribute to the gauge generator with coefficients dependent on the gauge parameters, alongside the first-class constraints; the answer is negative  \cite{ChitaiaGaugeFirstSecond}.  One thing that can happen is that one needs to redefine the constraints, mixing in some terms that naively look like second-class constraints into a modified set of first-class constraints.

However, once one sees how the Pooley-Wallace inverse Legendre transformation from the extended Maxwell Hamiltonian works to achieve an `extended' Lagrangian---one with a de-Ockhamized electrostatic potential $V$---it becomes clear how to make a massive generalization.  One can simply add a de-Ockhamized mass term to the Pooley-Wallace de-Ockhamized massless Lagrangian and then take the Legendre transformation to get the de-Ockhamized (`extended'?) massive Hamiltonian.  
Their de-Ockhamized  Lagrangian density (equation 48) is, after discarding an inessential spatial integration, 
$$ \mathcal{L}_{\mu^{\prime} }[\vec{A}, V; \dot{\vec{A}}, \cancel{ \dot{V}} ] =     \frac{1}{2} (\dot{\vec{A}} - \nabla(V + \mu^{\prime}) )^2     - \frac{1}{2}(\nabla \times \vec{A})^2 - ((V + \mu^{\prime}) \rho + \vec{A} \cdot \vec{J} ). 
$$
This is just the usual Maxwell Lagrangian density, but with $V + \mu^{\prime}$ playing the role of electrostatic potential.  To add a mass term one would usually add $- \frac{1}{2} m^2 ( -V^2 + \vec{A}^2 )$ to the Lagrangian density, but now we need to use the de-Ockhamized electrostatic potential, so the mass term is now 
$$ - \frac{1}{2} m^2 ( - (V + \mu^{\prime})^2 + \vec{A}^2).$$ 
Using the Pooley-Wallace expression plus the de-Ockhamized mass term, one can now perform the Legendre transformation in the usual way as far as possible (recalling that gauge freedom arises precisely because one cannot perform the full Legendre transformation). 
One gets for the canonical momenta
\begin{equation}  
 \vec{\pi}_E = \dot{\vec{A}} - \nabla (V + \mu^{\prime}),
\end{equation} 
which is equivalent to their equation 28, and 
$$\pi_0=0.$$
Calculating the total or primary Hamiltonian density (the one equivalent to the Lagrangian density, in this case equivalent to the de-Ockhamized Lagrangian density) yields, dropping a spatial divergence along the way,
\begin{eqnarray} 
\mathcal{H}_E^m = \frac{1}{2} \vec{\pi}^2 + \vec{\pi} \cdot \nabla (V + \mu^{\prime}) + \frac{1}{2} \vec{B}^2  + \frac{1}{2} m^2 \vec{A}^2  - \frac{1}{2} m^2 (V + \mu^{\prime})^2 + (V + \mu^{\prime}) \rho + \vec{A} \cdot \vec{J} + \lambda \pi_0  = \nonumber \\
\frac{1}{2} \vec{\pi}^2   + \frac{1}{2} \vec{B}^2     - (V + \mu^{\prime}) \nabla \cdot \vec{\pi}     -  m^2 (V + \mu^{\prime})^2 + \frac{1}{2} m^2 (V + \mu^{\prime})^2 + (V + \mu^{\prime}) \rho + \vec{A} \cdot \vec{J}+ \frac{1}{2} m^2 \vec{A}^2 + \lambda \pi_0  = \nonumber \\ 
\frac{1}{2} \vec{\pi}^2   + \frac{1}{2} \vec{B}^2     - [V + \mu^{\prime}) (\nabla \cdot \vec{\pi}    +  m^2 (V + \mu^{\prime})  -  \rho] + \vec{A} \cdot \vec{J}+  \frac{1}{2} m^2 (V + \mu^{\prime})^2 + \frac{1}{2} m^2 \vec{A}^2 + \lambda \pi_0.  
\end{eqnarray} 
For convenience, terms that turn out to be the secondary constraint have been gathered into square brackets.
This expression reduces to the usual Proca Hamiltonian given above for $\mu^{\prime}=0,$ as expected.  Perhaps less obvious is the fact that the de-Ockhamized Hamiltonian obtains simply from de-Ockhamizing the electrostatic potential; no terms involving the momenta are affected.  
While one does not normally talk of an extended Hamiltonian for a theory with no first-class constraints, such as Proca's massive electromagnetic theory, that is merely a verbal quibble.  The key feature of the `extended'  electromagnetic Lagrangian (Pooley and Wallace equation 48) is the de-Ockhamization of the electrostatic scalar potential---a process that one can perfectly well apply to the Proca Lagrangian.  
Requiring consistency over time by preservation of the primary constraint $\pi_0=0$ gives the secondary constraint
$$ \nabla \cdot \vec{\pi} + m^2(V + \mu^{\prime}) - \rho,$$ 
so the secondary constraint is de-Ockhamized \emph{via} the expression
 $$(V + \mu^{\prime}).$$
Preserving the de-Ockhamized secondary constraint with the de-Ockhamized Hamiltonian's time evolution gives
$$ - \nabla \cdot \vec{A}  + m^2 \lambda = 0,$$
which is the de-Ockhamized Lorentz condition, as one would expect  \cite[pp. 597-601]{Jackson}---a law, not just a good idea as in the massless case.  Thus no tertiary constraint arises.  
The primary constraint has zero Poisson bracket with itself.  The secondary constraint also has zero Poisson bracket with itself (not always a trivial result in field theory, because one is in fact dealing with infinitely many constraints, one per point, in contrast to particle theories, in which the antisymmetry of the Poisson bracket guarantees that result).  
But the cross-term is not zero: 
$$ \{ \pi_0(\vec{x}), \nabla \cdot \vec{\pi}(\vec{y}) + m^2 (V + \mu^{\prime})(\vec{y}) - \rho (\vec{y}) \}= - m^2 \delta(\vec{x}, \vec{y}).$$
Hence the constraints are second-class due to this cross-term.  This result is, however, a relation between the constraints, not a property of either one of them, a fact that will facilitate doing something unusual with the second-class primary constraint in massive electromagnetism.

The question of the sense in which the transformations generated by separately smeared constraints preserves the Hamiltonian field equations requires more scrutiny.  After presenting the equations for the extended Hamiltonian for Maxwell's (massless photon) electromagnetism, Pooley and Wallace inform us, let us recall, that ``[w]e can swiftly verify that the transformations generated by $O_t[f,g]$ are symmetries of \emph{these} equations for arbitrary $f$, $g$.''  While one actually cannot verify that swiftly (because it does not straightforwardly hold), the effort to do some of it slowly will be  rewarding.  Generalizing their equations (28-31) to admit the photon mass term, noting that simply setting $m=0$ recovers the massless case, one obtains: 
\begin{eqnarray} 
\dot{\vec{A}} = \frac{\delta H_E^m}{\delta \vec{\pi} } = \vec{\pi} + \nabla(V + \mu^{\prime}), \\
\dot{\vec{\pi}} = - \frac{\delta H_E^m}{\delta \vec{A} } =  - \nabla \times \vec{B} - \vec{J} - m^2 \vec{A}, \\ 
\dot{V} = \frac{\delta H_E^m}{\delta \pi_0} = \lambda, \\ 
\dot{\pi_0} = - \frac{\delta H_E^m}{\delta V} = \nabla \cdot \vec{\pi} - \rho + m^2(V + \mu^{\prime})=0.
\end{eqnarray} 
The primary constraint leaves the canonical momenta and $\vec{A}$ alone and changes the electrostatic potential $V$ by the arbitrary smearing function for both $m=0$ and $m \neq 0.$  The secondary constraint sprouts a new term for $m\neq 0,$ altering the canonical momentum $\pi_0$, which is bad, because that canonical momentum should stay $0$.  But one never expected anything gauge-like to arise from the second-class secondary constraint in massive electromagnetism, so this spoilage is  the kind of thing that one expected  \cite[pp. 183-186]{Sundermeyer}.  The effect of the primary constraint, namely shifting $V$ by the arbitrary smearing function $f$, leads to the following modified Hamiltonian field equations: 
\begin{eqnarray} 
\dot{\vec{A}}  =  \vec{\pi} + \nabla(V + \mu^{\prime} + f), \\
\dot{\vec{\pi}} =  - \nabla \times \vec{B} - \vec{J} - m^2 \vec{A}, \\ 
\dot{V} + \dot{f}  =  \lambda, \\ 
\dot{\pi_0} =  \nabla \cdot \vec{\pi} - \rho + m^2(V + \mu^{\prime} + f )=0.
\end{eqnarray} 
The first equation shows that the extended Hamiltonian sets a new, lower standard of having to further redefine the electrostatic potential by $f,$ rather than leaving the equation invariant as the standard Lagrangian-equivalent gauge transformations do.  Perhaps one can live with this lowered standard, as long as one implements it consistently.  The second equation is invariant.  The third equation involves redefining the arbitrary function $\lambda,$ which is not surprising given that this arbitrary function corresponds to the velocity of the electrostatic potential.  The fourth equation, however, contains a tragedy for the conventional view, because it shows that the \emph{ massive ($m\neq0$) equations are just as invariant under the transformation  generated by the primary constraint $\pi_0$ as the massless ($m = 0$) equations}:  one alters the de-Ockhamized electrostatic potential in the same way in equation 4 in the mass term as one does in equation 1 shared with the massless case.

  For friends of the extended Hamiltonian, the primary first-class constraint generates a gauge transformation for Maxwell's theory, suitably formulated with a de-Ockhamized electrostatic potential, but no second-class constraint is expected to generate a gauge transformation.   The new gauge freedom which pertains to the primary constraint alone, is, I say, pure artifice, having nothing essential to do with electromagnetic gauge freedom. That is the more obvious because  we now see that the primary \emph{second-class} constraint generates a gauge transformation for Proca's massive electromagnetism, suitably formulated with a de-Ockhamized electrostatic potential, by the same standard!  I don't think anyone thinks accepts such a result---neither the mathematically strict Hamiltonian-Lagrangian equivalent view  \cite{AndersonBergmann,CastellaniGaugeGenerator,PonsDirac,FirstClassNotGaugeEM} nor the conventional separate first-class constraints view \cite[pp. 19-21]{DiracLQM} holds that a second-class constraint generates a gauge transformation.  I take this to be a \emph{reductio} of the idea that de-Ockhamizing the electrostatic potential in electromagnetism introduces gauge freedom in any interesting sense.  In any case the conventional view that a first-class constraint (by itself) generates a gauge transformation (apart from a class of exceptions of no relevance here) \emph{and a second-class constraint does not generate a gauge transformation}, has a serious problem.  (Any view that omits the clause about second-class constraints is not orthodox.)  The orthodox are now logically committed to a heresy, whether by their own standards or the standards of the Lagrangian-equivalent reforming party.   One can preserve the phrase ``a first-class constraint generates a gauge transformation'' as a matter of theoretical choice (however doubtful its motivation apart from tradition, at least classically)---but one loses not only physical meaning, but also the accompanying claim (both mathematically strict and conventional)  that a second-class constraint does not generate a gauge transformation.  Hence it is quite misleading to claim to have vindicated the conventional claim that a first-class constraint generates a gauge transformation, because orthodoxy now entails a heresy that was not mentioned. The extended Hamiltonian finds gauge freedoms that aren't really there, because it installs them artificially by de-Ockhamizing, at least in electromagnetism; it therefore obscures rather than illuminates true gauge freedom by mixing counterfeits with the genuine article.  That is all the clearer now that we have managed to perform the analogous counterfeiting operation in massive electromagnetism, which naturally has no gauge freedom.\footnote{It is, of course, possible to install artificial gauge freedom in massive electromagnetism using an extra scalar field \cite{Ruegg,PittsArtificial,FrancoisArtificialSubstantialGauge}, a trick that can be very useful at times, both technically and conceptually. The resulting mass term is $- \frac{1}{2} m^2 (A_{\mu} + \partial_{\mu} \phi)(A^{\mu} + \partial^{\mu} \phi).$ One can, for example, show massive quantum electrodynamics to be renormalizable, or calculate the definition of observables given electromagnetic gauge freedom \cite{ObservablesEquivalentCQG}.}

The procedure employed to find (phony) gauge freedom in Proca's electromagnetism \emph{via} de-Ockhamization result differs in at least two key ways from Dirac's process of  extending the Hamiltonian for Maxwell's theory.  First,  only the primary (second-class) constraint, not the secondary (second-class) constraint, generates a gauge transformation (by the standards at hand).  Instead of Pooley and Wallace's $O[f,g] = \int (f C_0 - g C_1)$ that separately smears the primary and secondary constraints with arbitrary functions, I use an arbitrary smearing function $f$ for the primary, but set $g=0,$ omitting the secondary constraint from the game. A \emph{reductio} does not need to show that every second-class constraint generates a gauge transformation by the Pooley-Wallace criterion, but only that one of them does.  That is a good thing, because in fact only one of them, the primary constraint, actually does.  Second, one does not simply add terms in the secondary constraints to the Hamiltonian as in Dirac's extension procedure.  Instead one does whatever is required by de-Ockhamizing the electrostatic potential.  For the massless case, de-Ockhamization implies adding the secondary constraint(s) with an arbitrary coefficient(s); for the massive case it does not, because there are quadratic terms in the electrostatic potential to address.  

One might wonder why the second-class character of the primary constraint in Proca's massive electromagnetism does not spoil whatever `extended' gauge freedom might exist from the primary constraint in the massless theory.  The primary constraint the same in Proca's theory as in Maxwell's because the kinetic term, the piece of the Lagrangian with derivatives, is the same.  The primary constraint generates exactly the same transformation, adding an arbitrary smearing function to the electrostatic potential, whether or not there is a photon mass term.  Crucially, considered in \emph{isolation} there is nothing second-class about the primary constraint $\pi_0$:  its Poisson bracket with itself vanishes.  It is called second-class because of a \emph{relation} to the massive Gauss secondary  constraint.  But the massive Gauss constraint plays no role in this part of the de-Ockhamization process. Consequently it simply does not matter whether the primary constraint is first-class or second-class.  Thus  this trick manages to generate a gauge transformation from the second-class primary constraint in Proca by exactly the same standard that Pooley and Wallace employ in  finding a gauge transformation generated by the first-class primary constraint in Maxwell.

One can further understand why the second-class primary constraint in Proca's massive electromagnetism with de-Ockhamized electrostatic potential generates a gauge transformation in exactly the same sense as the first-class primary constraint Maxwell's massless electromagnetism as follows.  The real gauge transformations of the standard Maxwell Lagrangian density $- \frac{1}{4} F_{\mu\nu} F^{\mu\nu}$ are generated by the ``gauge generator'' $G$, a tuned sum of the secondary first-class constraint (phase space Gauss law) to modify $\vec{A}$ by a spatial gradient and the primary first-class constraint $\pi_0$, which changes the electrostatic potential by a time derivative of the same smearing function (up to a sign).  A photon mass term, if introduced, breaks this gauge freedom.  On the other hand, de-Ockhamization of the Maxwell Lagrangian introduces a new gauge freedom of sorts, generated by the primary constraint $\pi_0,$ which changes the electrostatic potential by itself in an arbitrary way.  As it happens, introducing a photon mass term and de-Ockhamizing the electrostatic potential do not get in each other's way.  So one can de-Ockhamize the electrostatic potential in massive electromagnetism and thus introduce a new gauge freedom of sorts, generated by the primary constraint $\pi_0$, which changes the electrostatic potential by itself in an arbitrary way.  The massive theory's second-class secondary constraint (modified phase space Gauss law, including a mass term bit) is not used.  Because  $\{ \pi_0(x), \pi_0(y) \}$ = 0, the primary constraint considered all by itself is, so to speak, first-class with itself.   For massive electromagnetism, the second-class nature of $\pi_0$ is not a property of $\pi_0,$ but a \emph{relation} with the secondary constraint (the Gauss law),   and that offending relatum is not used in a relevant way (as a generator) in the de-Ockhamization process.  Thus one sees why, if one accepts the conventional view that a first-class constraint generates a gauge transformation in Maxwell's electromagnetism, one is also committed to the apparently novel and heretical view that a second-class primary constraint generates a gauge transformation in Proca's electromagnetism.

The inverse Legendre transformation to a de-Ockhamized Lagrangian by Pooley and Wallace is thus a double-edged sword:  it simultaneously shows that  my argument against the empirical adequacy of the extended Hamiltonian formalism in section 9 of the 2014 \emph{Annals of Physics} paper was wrong, and shows the triviality of the gauge freedom generated by the primary constraint.  Indeed the recognition that de-Ockhamization is the engine powering the separate first-class constraints doctrine and the extended Hamiltonian  further opens the door (already foreshadowed in Pitts 2014) to seeing that a \emph{second-class} primary constraint generates a gauge transformation in Proca's massive electromagnetism in precisely the same way that a first-class primary constraint generates a gauge transformation in Maxwell's massless electromagnetism.  The massive generalization does not strictly involve adding a secondary constraint term (because another term arises due to the photon mass), but de-Ockhamizes the electrostatic scalar potential in exactly the same way.  That is a deeper  problem for `orthodoxy' than the unfortunate but isolated mistaken  section 9 of my 2014 paper.  It is still  either false or trivial that a first-class constraint (alone) generates a gauge transformation in Maxwell's electromagnetism and related theories such as Yang-Mills and General Relativity.  Hence the role of the separate-first-class-generates-gauge doctrine in setting up the problem of time in canonical General Relativity persists, and eliminating that doctrine still contributes much to resolving the problem of time \cite{GRChangeNoKilling}.  


%
%


\section{Response  to  Gomes and Butterfield} %

In a footnote Gomes and Butterfield have recently noted  that portions of  Henneaux and Teitelboim \cite[equations 19.11 and 19.13 with 3.26 and 3.31]{HenneauxTeitelboim} are relevant and, it is claimed, answer my arguments  \cite{GomesButterfieldMaxwellGauge}. Which arguments they have in mind is unclear in such a brief mention.  Charitably one might see them as criticizing especially my section 9, which claims to refute the extended Hamiltonian empirically; they do not notice that much of Pitts 2014 argues for ``false or trivial'' rather than ``false,'' but section 9 is the part of the paper that is successfully addressed by invoking this portion of Henneaux and Teitelboim.  Those portions of Henneaux and Teitelboim do not count, however, against the claim that it is either \emph{false or trivial}  that a first-class constraint generates a gauge transformation in electromagnetism and similar theories.  The awkward consequence that the second-class primary constraint generates a gauge transformation in Proca's massive electromagnetism by the same standard as the first-class primary constraint does in Maxwell's massless electromagnetism again weighs against efforts to rehabilitate the separate first-class constraints doctrine and the extended Hamiltonian. 


\section{Quantization?}

While the separate first-class constraints doctrine and extended Hamiltonian have little  to commend them classically, a quantum motivation is not necessarily excluded.  Henneaux and Teitelboim indeed offer a quantum motivation as one reason for postulating that first-class constraints typically generate gauge transformations.
\begin{quote} 
Third, as we shall see later, the known quantization methods for constrained systems put all first-class constraints on the same footing, \emph{i.e.}, treat all of them as gauge generators.  It is actually not clear if one can quantize at all otherwise. \cite[p. 18]{HenneauxTeitelboim}.  
\end{quote} 
Such an argument might be a good reason for adopting a formalism that has little to commend it classically. One also recalls that probably the first appearance of the idea that a first-class constraint generates a gauge transformation \cite{BergmannSchiller} also used quantum considerations.  

 On the other hand, the claim that key use is made of the separate first-class constraints doctrine is disputed by Pons in a section reflecting on why the defects in Dirac's work have not caused much trouble: 
\begin{quote} 
If Dirac's approach to gauge transformations was so incomplete, and his proposal
to modify the dynamics so gratuitous and unfounded, one may wonder to what
extent has it affected the correct development and applications of the theory. The
answer is: very little, and for various reasons.

First, because of the concept of observables, which will be the subject of the
following paragraph. Second, because Dirac's method of quantisation in the
operator formalism can be introduced either with the total Hamiltonian or with the
extended Hamiltonian with equivalent results. A third reason is that important
developments, for instance the most powerful theoretical tool for the quantization of
constrained systems, the field-antifield formalism (Batalin \& Vilkovisky 1981,
1983a,b; see Gomis, Paris, \& Samuel, 1995 for a general review), did not incorporate
these controversial features of Dirac's view. [A substantial footnote 11 discussing Henneaux and Teitelboim characterizes their approach as ``mixed.'']  Indeed, the natural concept of gauge
transformations---as Noether symmetries of the action---to be used in a path-integral
framework is that of mapping trajectories into trajectories (or field configurations
into field configurations).
\cite{PonsDirac} 
\end{quote} 
So Pons concludes that Henneaux and Teitelboim do not make essential use of the doctrine of separate first-class constraints as gauge generators after all.  Then quantization would not depend on the doctrine of gauge transformations from separate first-class constraints.  Clearly the truth of the matter is important for the purposes for which one studies gauge theories.


%
\end{document}